# INTERSTELLAR GRAINS: 50 YEARS ON

N.C. Wickramasinghe
Buckingham Centre for Astrobiology
The University of Buckingham
Buckingham MK18 1EG

Email: ncwick@gmail.com

## Abstract

Our understanding of the nature of interstellar grains has evolved considerably over the past half century with the present author and Fred Hoyle being intimately involved at several key stages of progress. The currently fashionable graphite-silicate-organic grain model has all its essential aspects unequivocally traceable to original peer-reviewed publications by the author and/or Fred Hoyle. The prevailing reluctance to accept these clear-cut priorities may be linked to our further work that argued for interstellar grains and organics to have a biological provenance – a position perceived as heretical. The biological model, however, continues to provide a powerful unifying hypothesis for a vast amount of otherwise disconnected and disparate astronomical data.

*If you can look into the seeds of time,
And say which grain will grow and which will not,
Speak then to me, who neither beg nor fear
Your favours nor your hate.....*

William Shakespeare:  *Macbeth  (*Act 1, Scene 3)

*\*\*\*\*\**
*Dedicated to Fred Hoyle (1915 – 2001)*

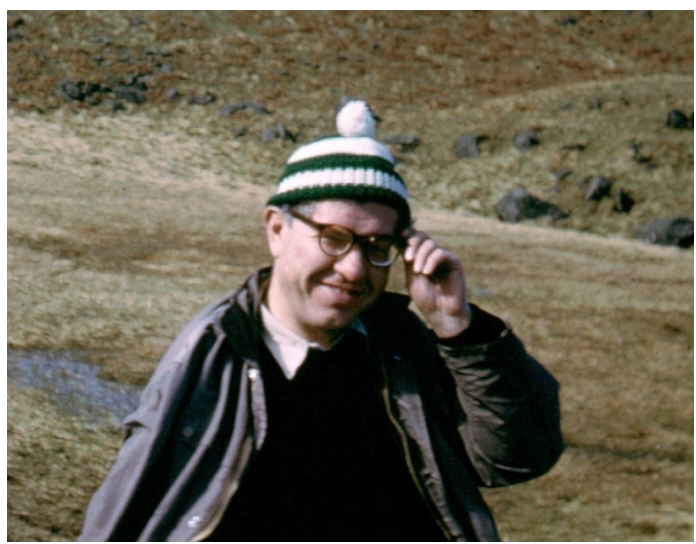

**Plate 1**.  Fred Hoyle (photographed here in 1961) and the author introduced and developed the theory of interstellar graphite particles in 1962, mixtures of graphite and silicate grains in 1969, organic grains in 1974 and biological grains in 1979.

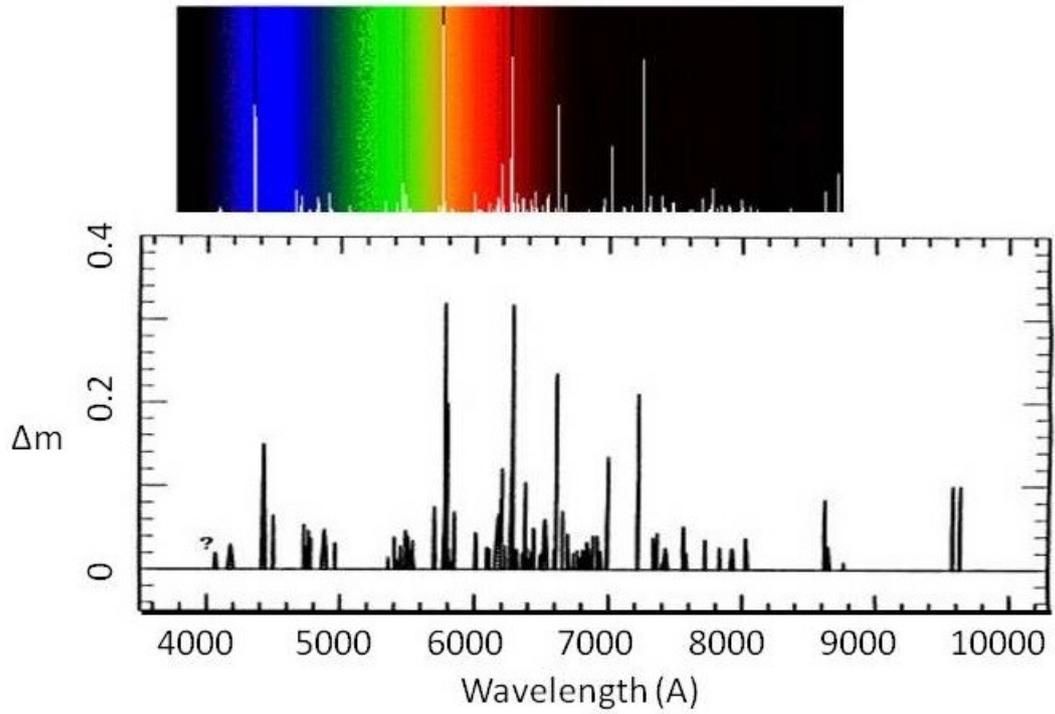

**Plate 2.** The diffuse interstellar bands with half widths in the range 2-30A are distributed over the entire visual waveband. They are associated with interstellar dust grains but have defied identification for nearly 100 years.

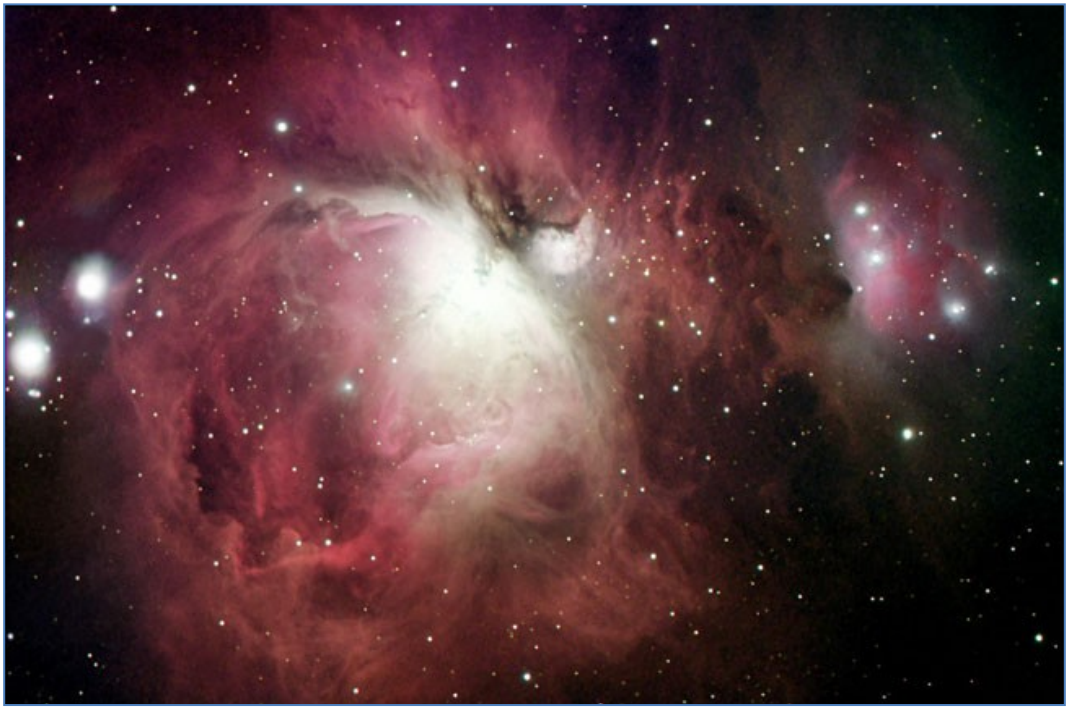

**Plate 3**. The Orion nebula the birthplace of stars and planets

## 1. Introduction

The present writer commenced his studies on interstellar dust precisely 50 years ago. In half a century the subject has grown immensely: a relatively obscure field of astronomy studied by a handful of researchers in 1961 now engages the attention of many thousands of investigators. Several satellites and space telescopes are dedicated to the study of interstellar dust, and others are expected to be operational in the near future.

What precisely are interstellar grains made of? How are they formed, and what is their role in the formation of stars, planets – and life? After 50 years the answers to these questions are far from settled.

In the autumn of 1961, Fred Hoyle and the author embarked on a scientific journey to understand the nature of interstellar dust. At the time the firmly held belief was that interstellar grains were dirty ice particles that nucleated and grew in the tenuous clouds of interstellar space. This model was soon shown by us to be untenable and we were led to propose the alternative graphite particle theory (Hoyle and Wickramasinghe, 1962). The emphasis then shifted sharply from the formation of icy grains in interstellar clouds (Oort and van de Hulst, 1946) to condensation of refractory particles in the mass flows from cool stars (Wickramasinghe, 1967). Over the next two decades Hoyle and the present writer proposed a succession of refinements to our original model that were dictated by new astronomical data. These developments are set out in tabular form below (Table 1):

Table 1
Chronology of salient developments, each contribution representing the very first publication on a particular topic

| | | | |
|---|---|---|---|
| a | 1962 | Graphite particle theory; formation of grains in stellar mass flows | *MNRAS*, **124**, 417-433 |
| b | 1963 | Optics of graphite grains | *MNRAS*, **126**, 99-114 |
| c | 1965 | Core-mantle grains: ice mantle growth on graphite grains | *MNRAS*, **131**, 177-190 |
| d | 1969 | Graphite silicate grain mixtures and interstellar extinction curves | *Nature*, 223, 450-462 |
| e | 1974 | Organic polymers in the interstellar medium - polyformaldehyde | *Nature,* **252**, 462-463 |
| f | 1977 | Aromatic molecules and the 2200A interstellar band | *Nature*, **270**, 323-324 |
| g | 1977 | Polysaccharides and the IR spectra of galactic sources | *Nature*, **268**, 610-612 |
| h | 1977 | Prebiotic polymers in space | *Nature*, **269**, 674-676 |
| i | 1979 | Desiccated bacterial grains and the optical extinction curve | *ApSS*, **66**, 77-90 |
| j | 1982 | GC-IRS7 and the infrared spectrum of dry bacteria | *ApSS*, **83**, 405-409 |
| k | 1986 | IR spectrum of Halley's comet and the bacterial dust model | *Earth, Moon, and Planets*, **36**, 295-299 |

No single step in this sequence was trivial nor was it taken lightly. The progression from step (g) to step (k), however, was particularly fraught with problems – problems that were more connected with sociology than science.

The idea that a significant mass fraction of interstellar dust is biologically generated was considered heresy. But to the annoyance of very many colleagues we argued in 1979 that biology was a unifying hypothesis for a large body of astronomical data (Hoyle and Wickramasinghe, 1979a,b; Hoyle et al, 1982a,b). The price to be paid for our obstinacy was that even our earlier innovative contributions in the chain of logic that led to the "heresy" came to be ignored. No reference to us is made in the adoption of the graphite particle model of grains – one for which we fought hard at conferences during the period 1962-1967. The currently popular MRN model of graphite-silicate mixtures (Mathis et al, 1977) makes no reference to the original paper in *Nature* (Hoyle and Wickramasinghe, 1969) in which precisely the same grain mixtures and extinction curves were discussed. This trend continues to the present day. In a recent review article (Draine 2003) an extensive treatment of extinction by graphite-silicate mixtures is given without any reference to the first publications on this topic (Hoyle and Wickramasinghe, 1969, 1991). Scientific etiquette is cast to the winds in an attempt to disinherit us from the publication priorities that are unequivocally ours!

In 1986 new observations of Comet Halley led to the discovery of cometary dust being spectroscopically similar to interstellar dust and to biological material. We had in fact predicted the spectrum of cometary dust precisely as it was observed. The lack of any attributions to us led to a correspondence in the columns of *Nature* (Hoyle and Wickramasinghe, 1982) to which John Maddox, Editor of *Nature*, replied with a vituperative piece entitled: "When reference means deference" – and deference we were surely denied (Maddox, 1986).

## 2. Convergence to biology

Notwithstanding such sociological obstacles, convergence towards correct ideas in science proceeds inexorably. The observational situation in 2011 is dramatically different from that which prevailed when the author started his journey in 1961. Organic molecules are everywhere: organic dust is all-pervasive; PAH's occupy every corner of the universe (Smith et al, 2007; Kwok, 2009; Rauf and Wickramasinghe, 2010). How are such materials formed and what is their significance? On the Earth over 99.999% of all the organic material is unquestionably biogenic. Why is it not reasonable to explore the same option for astronomy? Being forbidden by convention is not a good enough reason.

Biology on a cosmic scale is considered by some as an "extraordinary hypothesis" and it is stated that extraordinary evidence is needed to defend it. On the contrary confining life to Earth could be regarded as a far more extraordinary assertion, so it is the defence of this latter point of view that requires extraordinary evidence! And such evidence is of course non-existent, or at best illusory.

The overriding justification for grains, or a significant fraction thereof, to be somehow connected with biology stems from the argument that life itself could *only* have arisen in a cosmological setting. Probability arguments demand a setting for an origin of life that transcends enormously the miniscule scale of our planet (Hoyle and Wickramasinghe, 2000). Such a cosmological setting has recently been discussed by Gibson, Schild and the present author (Gibson et al, 2010). A cosmologically derived legacy of life along with its full range of evolutionary potential (bacteria and viruses) were introduced *via* frozen comets and planets into galaxies such as our Milky Way system (Napier et al, 2007; Wickramasinghe, J. et al, 2010; Gibson et al, 2010). Microbial life is thereafter amplified and recycled between billions of planetary abodes, of which our solar system is just one. Microbial material, on this picture, must escape continuously into the interstellar medium from comets and planetary systems. A large fraction of the "PAH's" and other organic molecules discovered in the galaxy could represent biological material in various stages of degradation.

The author's conviction of the correctness of this approach emerged with step (j) in the progression listed in Table1. The first mid-infrared observations of GC-IRS7 near the galactic centre revealed the average properties of interstellar dust over a distance scale of some 10kpc (Allen and D.T. Wickramasinghe, 1981). Combined with the already available extinction curves in the visual and ultraviolet spectral regions we arrived at the correspondences for the bacterial grain model with astronomical data shown in the curves of Fig. 1. Both the 2175A ultraviolet extinction peak assigned to biological aromatics (Hoyle and Wickramasinghe, 1977) and the 3 - 4μm feature due to various CH stretching modes (Hoyle et al, 1982) stood out like a pair of beacons reaffirming the validity of the theory of cosmic life. So it seemed to us in 1982, and this conviction continues to grow. In subsequent sections we explore further a few more aspects of astronomical observations that are unified by the biological hypothesis.

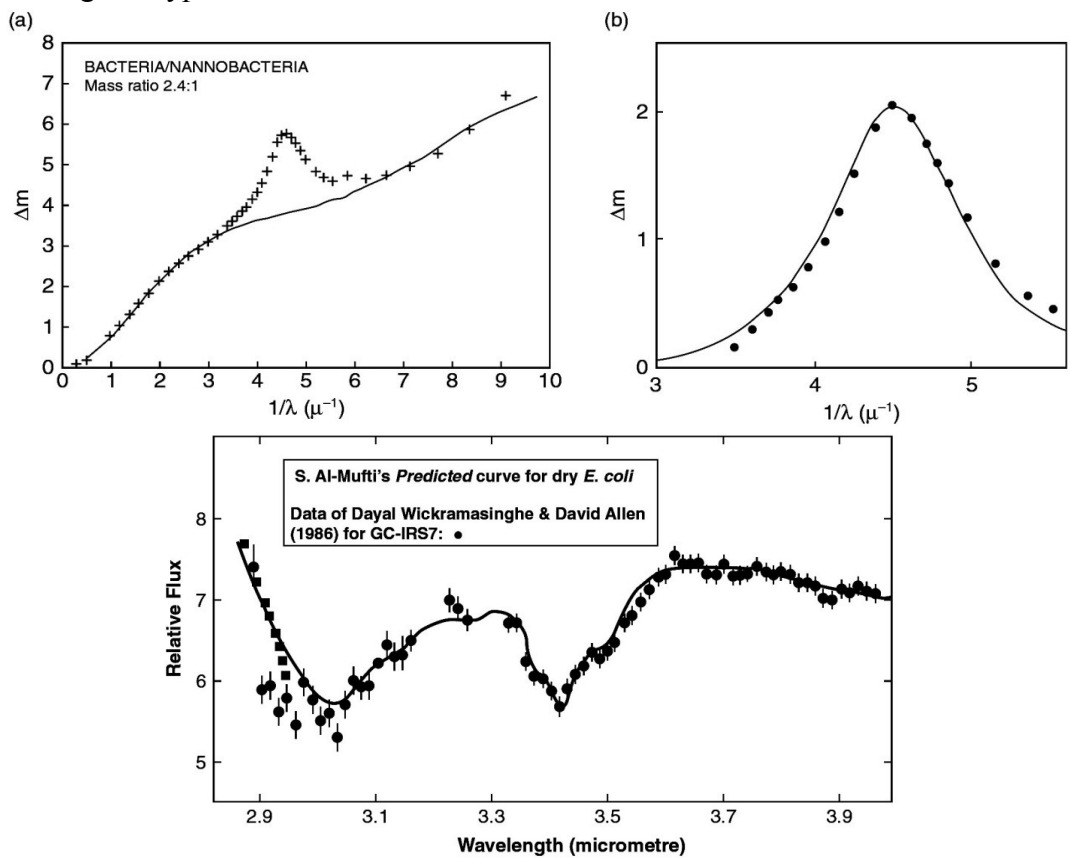

Fig. 1 *Upper* (a) The mean extinction curve of the galaxy (points) compared with the contribution of desiccated bacteria and nanobacteria.
*Upper* (b) The residual extinction compared with the normalized absorption coefficient of an
ensemble of 115 biological aromatic molecules.
*Lower:* The first detailed observations of the Galactic centre infrared source GC-IRS7 (Allen & Wickramasinghe 1981) compared with earlier laboratory spectral data for dehydrated bacteria.

We conclude this section with a montage of 3-3.8 μm spectra of astronomical sources (including comets) that can be fitted to spectra of coals in various stages of degradation. Coals are of course degradation products of biomaterial. The structural formulae in (c) are inferred from mass spectroscopy of interstellar dust studied using equipment on STARDUST (Kreuger et al, 2004).

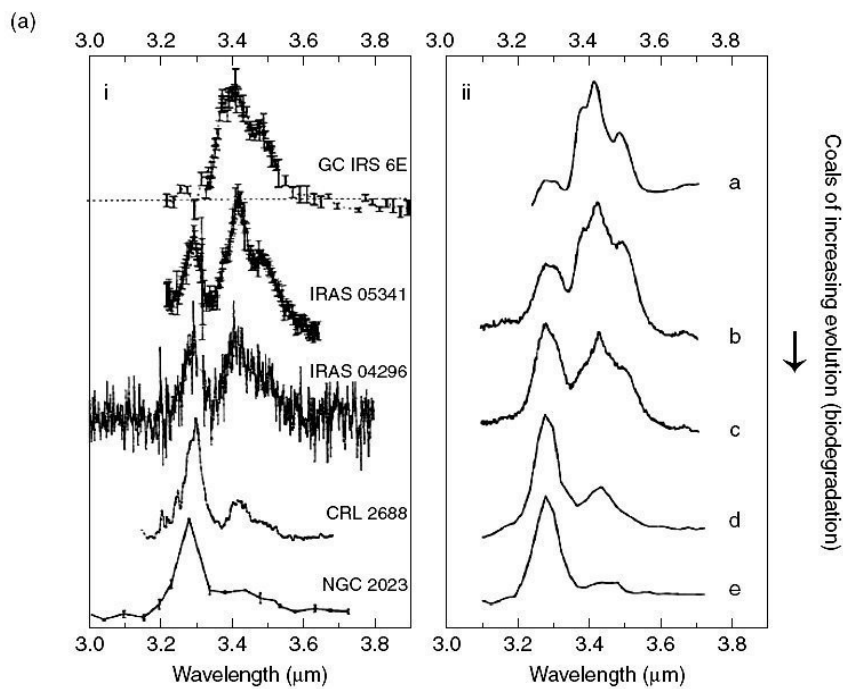

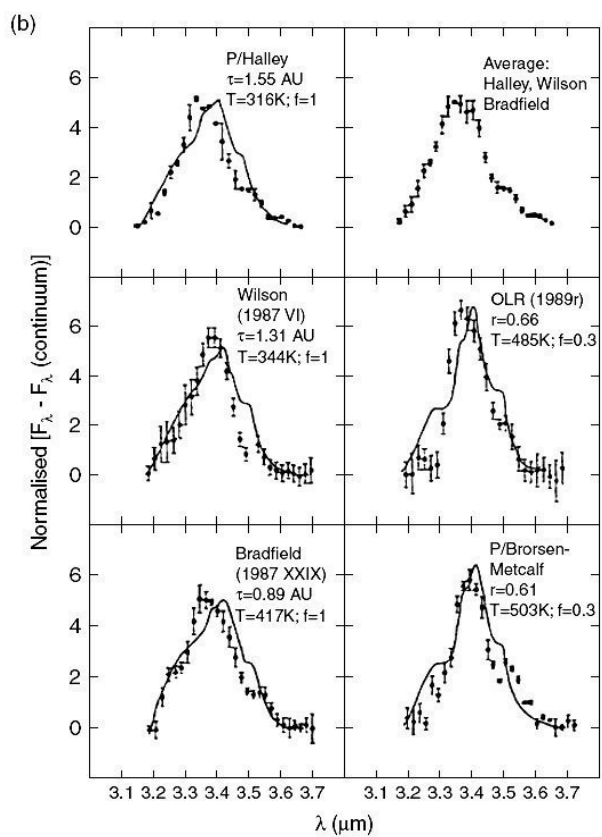

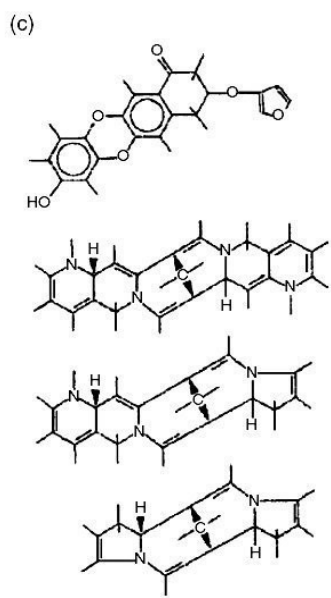

Fig. 3 Normalized absorption profiles of a number of astronomical infrared sources compared with spectra of coals of varying degrees of degradation – (i) being the closest to desiccated bacteria. (b) The points represent the 3.1–3.8 mm emission profiles of several comets. The curves are for bacterial PAH where f is the ratio of opacities arising from aromatic molecules at 3.28 mm to that from E-coli at 3.4 mm. (c) Functional groups in the break-up fragments of impacting interstellar dust grains, inferred by Krueger et al (2004) from mass spectroscopy.

## 3  Extended red emission and PAH-related biomolecules

The detection of an extended red emission (ERE) over the waveband 6000-8000A in planetary nebulae (Witt and Schild, 1985; Witt, Schild and Kraiman, 1984; Furton and Witt, 1990) could now be interpreted as evidence for the presence of fragmented biomaterial in these objects. Since its original discovery ERE has been observed in a wide variety of dusty regions in our galaxy (Perrin and Sivan, 1992) and in external galaxies as well (See review by Hoyle and Wickramasinghe, 1996). ERE has also been observed in the diffuse interstellar medium (Gordon et al, 1998) and in high latitude galactic cirrus clouds (Witt et al, 2008). Although it is widely held that ERE is caused by simple (possibly compact) PAH's under a variety of excitation conditions, fits to astronomical data leaves much to be desired. On the other hand fluorescence in fragments of biomaterial such as chloroplasts offers a better prospect, and these could play a role in explaining the entire set of astronomical observations. Fig. 2 compares the fluorescence behaviour of fragmented spinach chloroplasts (Boardman et al, 1966) with the observations for a planetary nebula NGC7027 (Furton and Witt, 1990). The general agreement is seen to be satisfactory.

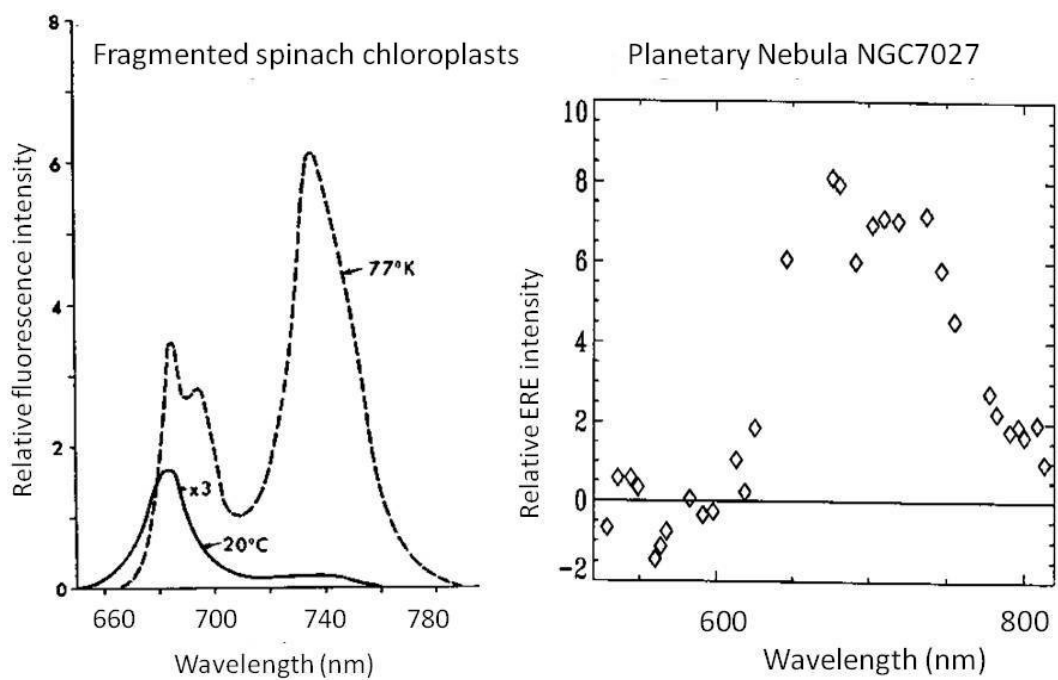

Fig. 2 Spectra of fragmented spinach chloroplasts at two temperatures (Boardman et al, 1966) and spectra of ERE excess in NGC7027

Chloroplasts are not presented here as a defintive or unique identification of the ERE carrier, but merely as an illustration of the types of PAH's associated with biological fragments that could collectively fit the astronomical data better than abiotic PAH's. It is also worth noting that the biological structures that give rise to ERE may also be responsible for many of the observed unidentified infrared bands (UIB's) (Smith et al, 2007; Thaddeus, 2006; Kwok, 2009; Kwok and Zhang, 2011) and indeed for the 2175A absorption band (Hoyle and Wickramasinghe, 1977). Table 2 lists the absorption wavelengths of several types of biological systems that show correspondences with the observed UIB bands (Rauf and Wickramasinghe, 2010).

Table 2. Distribution of two astronomical observations (UIBs and proto-planetary nebulae (PPNe)) and major IR absorption bands in laboratory models of terrestrial origin

| UIBs | PPNe | Algae | Grasses | Bituminous coal | Anthracite coal |
|---|---|---|---|---|---|
| 3.3 | 3.3 | 3.3 | – | 3.3 | 3.3 |
| – | 3.4 | 3.4 | 3.4 | 3.4 | 3.4 |
| 6.2 | 6.2 | 6.0 | 6.1 | 6.2 | 6.2 |
| – | 6.9 | 6.9 | 6.9 | 6.9 | 6.9 |
| – | 7.2 | 7.2 | 7.2 | 7.2 | 7.2 |
| 7.7 | 7.7 | – | 7.6 | – | 7.7 |
| – | 8.0 | 8.0 | 8.0 | – | – |
| 8.6 | 8.6 | 8.6 | – | – | – |
| 11.3 | 11.3 | 11.3 | 11.1 | 11.5 | 11.3 |
| – | 12.2 | 12.1 | 12.05 | 12.3 | 12.5 |
| – | 13.3 | – | – | – | 13.4 |

### 3 Diffuse interstellar bands and the visual interstellar extinction

A set of astronomical data that has not been explained in nearly 100 years are the diffuse interstellar absorption bands in the optical spectra of stars. The strongest of these is centred at 4430A and has a half-width of ~ 30A. A possible solution to a 100 year old unsolved problem may also be connected with the behaviour of fragmentation products of biology existing under various conditions of excitation in their electronic configurations. A possible candidate in this category was proposed by F.M. Johnson in the form of magnesium tetrabenzo porphyrine (Johnson, 2006).

Since the strength of the 4430A band is known to be tightly correlated to the colour excess $E_{B-V}$ and thus to the total extinction, the possibility that the continuum extinction itself has a significant contribution arising from molecular absorptions merits serious consideration. The points in Fig 4 shows the average extinction data for several thousand stars in the optical spectral region studied by Nandy (1964, 1965), which represents perhaps the most comprehensive dataset available over this waveband. The striking feature of Fig. 4 is the strictly linear behaviour of this curve ($\Delta m$ *vs* $1/\lambda$ over the range $1.2 \leq \lambda^{-1} \leq 2.4 \mu m^{-1}$ with a sharp change in slope at $\lambda^{-1} = 2.4 \mu m^{-1}$   A later study by Schild (1977) confirms this result whilst pointing out a third change of linear slope at $\lambda$ = 3200A. At shorter wavelengths the extinction curves cease to be invariant with varying strengths of the $\lambda$ 2175A feature.

For particles with refractive index n in the range 1.33-1.66 (ices and silicates), fitting this extinction curve to models requires arbitrary fine-tuning of the mean size and size distribution to a level of precision that cannot be justified (see for instance the curves in Wickramasinghe, 1973). The arbitrariness of the choices is unacceptable in view of the almost precise invariance of this segment of the extinction in all directions and in all sources. Hoyle and Wickramasinghe (1977) attempted to solve this problem by invoking porous or hollow grains with $<n>$ = 1.167, for which the size constraints are significantly relaxed and the result is shown in the solid curve of Fig.4.

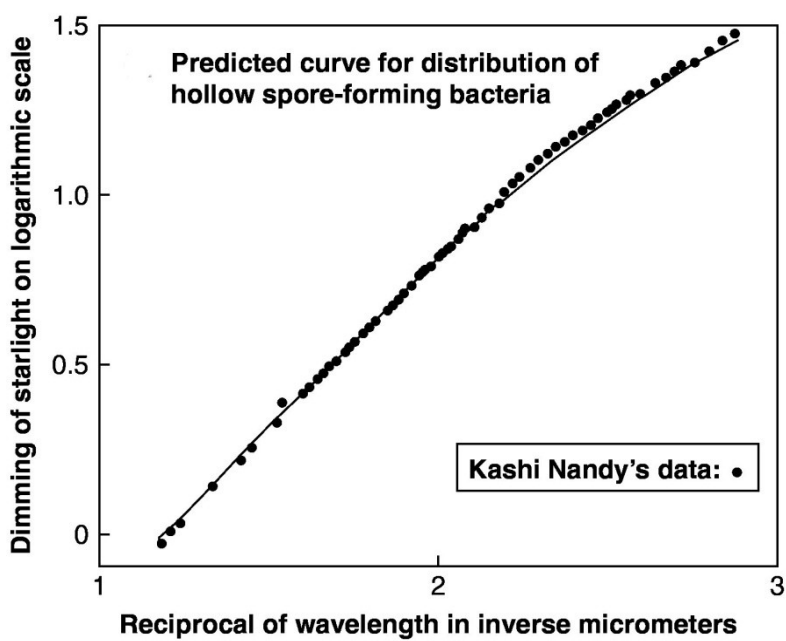

Fig. 4  Points represent the visual extinction data normalized to $\Delta m$=0.409 at $1/\lambda$=1.62 $\mu m^{-1}$ and $\Delta m$=0.726 at $1/\lambda$=1.94 $\mu m^{-1}$ (Nandy, 1964); the curve is the calculated extinction curve for a size distribution of freeze-dried spore-forming bacteria. The calculation uses the classical Mie theory and assumes hollow bacterial grains comprised of organic material with refractive index n=1.4 and with 60% vacuum cavity caused by the removal of free water under space conditions.

### 4   A revival of Platt particles

A molecular absorption model for the visual extinction curve was originally proposed by Platt (1956). Since 1956 much experimental work has been done on spectroscopy of organic dye molecules, for instance in relation to their role in enhancing the light gathering efficiency of solar cells (Snaith et al 2008). It should be noted, however, that spectra of dye molecules in solution or adsorbed on surfaces have very much broader absorption bands compared with free gaseous molecules.

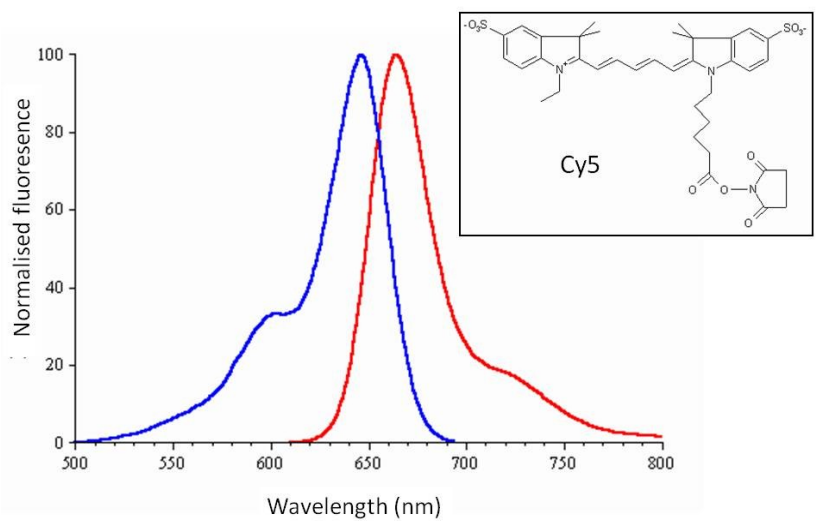

Fig. 5 Excitation spectrum (blue) and emission spectrum of Cy5 cyanine dye (Inset is structure of the dye molecule

Fig.5 shows the excitation/emission spectra of Cy5, a molecule used extensively in confocal microscopy. The set of molecules Cy2, Cy3, Cy5 defines a monotonically increasing length sequence that corresponds to a monotonic increase in absorption/emission wavelength. (For contributing to the extinction of starlight molecular absorption is all that is relevant; re-emission occurs in a random direction and does not therefore contribute to removal of energy from the forward beam.) For a dye molecule with a linear chain of conjugated bonds the peak of visible absorption is due to delocalised π-electrons. To a good approximation the system can be treated as a free electron (particle in a box) model amenable to description by Schroedinger's equation. The length of the box L is effectively the length of the chain of conjugated bonds and the main permitted transitions are from the highest occupied energy level to the next. In such a simplified model it can be shown that the emission peak is at the wavelength

$$\lambda \cong \frac{8mL^2 c}{(N+1)h}$$

where N is the number of free electrons in the conjugated chain and L is the length of the box. With N proportional to length L we obtain

$$\lambda \approx \text{constant.} L$$

Platt's original analysis of dye molecules yielded a constant equal to about ~ 400, giving λ ≈ 6000A for a 15A long chain. A Platt particle model might provide an explanation of the invariance of the visual extinction curve discussed earlier, provided an appropriate length distribution of organic dye molecules could be justified.

## 5  Size-distribution of biological fragments

Bacterial particles injected into interstellar clouds would suffer desiccation, and a large fraction could also undergo explosive fragmentation. A desiccated bacterium is expected to be relatively fragile in view of the fractal nature of macro-molecular assemblages. Sporadic spikes in the net electric charge on a grain – e.g. due to cosmic ray interaction could lead to "explosion" or break-up whenever the electromagnetic stress ~ $E^2/8\pi$ exceeds the binding between fractal units. Similar effects are well documented during transport of powders where explosions are due to frictional charging of small particles (Glor, 2010).

Carpinteri and Pugno (2002) have derived "universal laws" for such fragmentation and explosions. For a 3-dimensional structure the probability density of fragments with radii between r and r+dr is given as

$$p(r)dr = D \, \frac{r_{min}^D}{r^{D+1}} dr \qquad (1)$$

where $r_{min}$ is the minimum fragment size and D is the fractal dimension. A fractal dimension D ~ 2 would seem to hold for a wide range of biological assemblages such as viruses, and thus we set D=2 in the discussion that follows. Then we have a size distribution of fragments

$$n(r)dr \propto r^{-3} dr \qquad (2)$$

For a set of organic dye-like molecules we assume that the peak absorption wavelength λ is defined by

$$\lambda_m = Cr, \quad C \cong 400 \qquad (3)$$

as proposed by Platt (1957). Furthermore, we assume that the effective cross-section for absorption is $\sigma \propto r^2$ and that its wavelength dependence is defined to a good approximation by a delta function peaking at $\lambda = \lambda_m$. Thus we have a cross-section

$$\sigma(r,\lambda) \propto r^2 \delta(\lambda - Cr) \qquad (4)$$

where $\delta(x)$ is the Dirac delta function.

The mean extinction cross-section of starlight now requires an integration over r, with the size distribution of dye fragments as a weighting factor. From (2) and (4) we obtain

$$\Delta m(\lambda) \propto \int_0^\infty \sigma(r,\lambda) \, n(r) dr$$
$$= \int_0^\infty r^2 \delta(\lambda - Cr) r^{-3} \, dr$$
$$= const. \, 1/\lambda$$

thus providing the required invariant wavelength dependence of visual extinction.

The extinction curve depicted in Fig. 4 that we earlier fitted to desiccated bacteria (solid curve) would have an alternative explanation in terms of a molecular absorption model. At any rate an initial extinction curve arising primarily from in tact desiccated bacteria would not be perturbed by the presence of disintegration products that also contribute to $1/\lambda$ extinction. It has to be admitted that the analysis leading to this statement needs to be backed up with experimental data relating to the fragmentation of microorganisms. Laboratory studies that are needed to confirm both the size spectrum of fragmentation obtained from the Carpenteri-Pugno theory, as well as the linear relation between the wavelength of maximum absorption and length given in equation (3). Until such time the conclusions reached in this section must be regarded a provisional.

## 6  Concluding remarks

In five decades of research into the nature of interstellar grains there have been many unexpected changes of paradigm. Van de Hulst's ice grain theory seemed set in stone when the author first began his studies in 1961, but its eventual overthrow did indeed occur. The introduction of the graphite particle theory, initially vigorously opposed, gained ground over a decade, following which its proponents felt obliged to abandon it in favour of polymeric grains and aromatic molecules in 1977 (Wickramasinghe, 1974; Hoyle and Wickramasinghe, 1977). Silicate grains came rapidly into vogue from 1969 onwards (Hoyle and Wickramasinghe, 1969). However pinning down a particular silicate proved so difficult that astronomers felt obliged to postulate the existence of an "astronomical silicate" – thus inverting the problem and defining a hypothetical absorbing material to match astronomical spectra over the 8-12 μm waveband. Although the presence of silicate particles in interstellar space cannot be denied (we are standing on a silicate planet!), the overwhelming dominance of organics and organic polymer over the wavebands 3-4 μm, 8-14 μm and 18-22 μm has been reaffirmed over many years (Hoyle and Wickramasinghe, 1991).

The universe is surely far stranger that we could ever imagine, and an endless series of further surprises must lie in store. The 1970's, 1980's and 1990's witnessed the unravelling of a veritable Pandora's box of organics in every nook and cranny of the cosmos. With a large fraction of all the carbon in the universe locked up as organic, life-like material the connection with life stares one in the face. Two choices remain. Are we to suppose that we are witnessing an incredibly difficult, near impossible, progression from non-life to life occurring everywhere? Or could we be seeing evidence of the omnipresence of cosmic life and the inevitability of panspermia? The present author opted for the latter – life being a cosmic phenomenon. But the reader might go for the former as being socially more acceptable in the year 2011.

The biological theory of grains, admittedly unorthodox at the present time, has the merit of serving as an unifying hypothesis for an incredibly vast range of observations. Alternative non-biological explanations appear to be highly contrived, demanding fine-tuning of large numbers of free parameters. The situation strikingly reminiscent of the Ptolemaic model of the Solar System half a millennium ago – a new epicycle being required for every new observation.

With a few exceptions astrobiologists have chosen to disconnect astronomical phenomena from biology. The trend is to interpret a vast body of evidence that now exists for biochemicals in interstellar space as evidence of a *hypothetical* prebiotic chemistry operating on an astronomical scale. Such a presumption has no factual basis whatsoever, and it is likely to be woefully wrong. It is far more probable that the origin of life was a unique event that happened only once in the very early history of the Universe (Gibson, Schild, Wickramasinghe, 2010). The degradation of living cells is a well-understood process: the transformation of organic life on the Earth through a series of steps leading to anthracite and coal is well documented. The astronomical data discussed in this paper – the 2175A absorption, the unidentified infrared and visual bands and ERE - have a far better chance of being correctly explained as evidence of post-biology rather than prebiology. However, the last word has yet to be said.

In conclusion it should be stressed that the twin problems of the origin of life and the nature of interstellar grains remain far from solved at the present time. It would be prudent to preserve a sense of history and of humility in relation to all such big questions in Science. In 2011 brand new vistas are opening in front of us, and it would be future generations that would have the privilege to explore them.